\documentclass[
reprint,
superscriptaddress,
amsmath,
amssymb,
aps,
pra,
]{revtex4-2}

\usepackage{graphicx}
\usepackage{dcolumn}
\usepackage{bm}
\usepackage[colorlinks=true,linkcolor=blue,citecolor=blue,urlcolor=blue]{hyperref}

\usepackage{xcolor}
\usepackage{overpic}
\usepackage{comment}
\usepackage{dsfont}
\usepackage{physics}

\usepackage{cleveref}
\crefname{equation}{Eq.}{Eqs.}
\Crefname{equation}{Equation}{Equations}
\crefname{figure}{Fig.}{Figs.}
\Crefname{figure}{Figure}{Figures}
\crefname{section}{Sec.}{Secs.}
\Crefname{section}{Section}{Sections}
\crefname{appendix}{Appendix}{Appendices}
\Crefname{appendix}{Appendix}{Appendices}

\renewcommand{\d}{\mathrm{d}}
\newcommand{\ee}{\mathrm{e}}
\newcommand{\ii}{\mathrm{i}}

\begin{document}
	\title{Mediated Transmission of Quantum Synchronization in Star Networks}
	
	\author{Shuo Dai}
	\email{deus99@ruc.edu.cn}
	\affiliation{School of Physics, Renmin University of China, Beijing, 100872, China}
	\author{Ran Qi}
	\email{qiran@ruc.edu.cn}
	\affiliation{School of Physics, Renmin University of China, Beijing, 100872, China}
	\affiliation{Key Laboratory of Quantum State Construction and Manipulation (Ministry of Education), Renmin University of China, Beijing, 100872, China}

	
	\begin{abstract}
		Synchronization transmission describes the emergence of coherence between two uncoupled oscillators 
		mediated by their mutual coupling to an intermediate one. 
		In classical star networks, such mediated coupling gives rise to remote synchronization—%
		where nonadjacent leaf nodes synchronize through a nonsynchronous hub—%
		and to explosive synchronization, characterized by an abrupt collective transition to coherence. 
		In the quantum regime, analogous effects can arise from the interplay between 
		$1\!:\!1$ phase locking and $2\!:\!1$ phase-locking blockade in coupled spin-1 particles. 
		In this work, we investigate a star network composed of spin-1 particles. 
		For identical oscillators, symmetric and asymmetric dissipation lead to distinct transmission behaviors: 
		remote synchronization and quasi-explosive synchronization appear in different coupling regimes, 
		a phenomenon absent in classical counterparts. 
		For nonidentical networks, we find that at large detuning 
		remote synchronization emerges in the weak-coupling regime 
		and evolves into quasi-explosive synchronization as the coupling increases, 
		consistent with classical star-network dynamics. 
		These findings reveal the rich dynamical characteristics of mediated quantum synchronization 
		and point toward new possibilities for exploring synchronization transmission 
		in larger and more complex quantum systems.
	\end{abstract}

	\maketitle

	\noindent

	\section{Introduction}
	Synchronization is a fundamental phenomenon in nonlinear science, 
	describing the spontaneous adjustment of frequencies among interacting oscillators 
	toward a common value~\cite{pikovsky2001universal,boccaletti2002synchronization,acebron2005kuramoto}. 
	It appears ubiquitously across natural and engineered systems, 
	spanning biological, chemical, electronic, and mechanical oscillators~\cite{wiesenfeld1996synchronization,kiss2002emerging,hoppensteadt1999oscillatory,pecora1990synchronization,rulkov1996images,schafer1998heartbeat,goldstein2009noise,fell2011role,matheny2014phase}. 
	In classical systems, synchronization arises through the mutual coupling 
	of limit-cycle oscillators, leading to phase locking 
	and collective entrainment phenomena that are well described 
	within the framework of nonlinear dynamics and the Kuramoto model~\cite{kuramoto2005self}. 
	These studies have established synchronization as a cornerstone 
	of collective behavior in driven-dissipative systems, 
	laying the foundation for exploring its manifestations 
	in more complex network topologies and quantum regimes.

	The synchronization behavior of quantum harmonic oscillators 
	has been widely explored in recent years~\cite{lee2013quantum,walter2014quantum,lee2014entanglement,walter2015quantum,dutta2019critical,sonar2018squeezing}. 
	Beyond these continuous-variable systems, 
	the synchronization of few-level quantum oscillators 
	with incoherent gain and damping has attracted growing attention, 
	as it captures essential features of quantum phase locking 
	in open-system settings~\cite{roulet2018synchronizing,roulet2018quantum,schmolke2022noise,kehrer2024quantum,parra2020synchronization}.
	These systems encompass both the synchronization of a single spin 
	to an external signal and mutual phase locking between coupled spins. 
	Experimentally, quantum synchronization has been demonstrated 
	across a variety of physical platforms, 
	including trapped ions~\cite{zhang2023quantum}, 
	superconducting circuits~\cite{koppenhofer2020quantum}, 
	and cavity-QED systems with solid-state or cold-atom implementations~\cite{bagheri2013photonic,laskar2020observation}.

	In a three-level quantum system, a dissipative configuration 
	in which both the upper and lower states decay into the intermediate state 
	can generate a stable limit cycle~\cite{roulet2018synchronizing}. 
	When projected onto the Bloch sphere, the steady-state distribution 
	forms a uniform ring along the equator, 
	representing oscillations in the azimuthal direction. 
	When two spin-1 particles are coupled, they tend to align their phases, 
	and a stable $1\!:\!1$ phase locking emerges when their relative phase converges to a single value, 
	indicating synchronized dynamics. 
	In contrast, if the phase alignment occurs at two distinct relative angles, 
	the $1\!:\!1$ phase locking is suppressed and replaced by a $2\!:\!1$ phase locking, 
	corresponding to the interference-blockade regime. 
	Such interference blockade arises naturally between two identical coupled spin-1 particles~\cite{roulet2018quantum}. 
	Moreover, two identical spin-1 particles that are not directly coupled 
	can become synchronized through an intermediate identical spin-1 particle, 
	which mediates a $1\!:\!1$ phase-locking channel between them
	when all three oscillators share symmetric dissipation~\cite{kehrer2024quantum}.
	
	In this work, we investigate synchronization transmission 
	in a star network composed of spin-1 particles. 
	Remote synchronization, where the leaves become phase locked through an unsynchronized and detuned hub, 
	can arise in classical star networks~\cite{PhysRevE.85.026208}, 
	and increasing the coupling strength can further lead to quasi-explosive synchronization~\cite{gomez2011explosive}, marked by a rapid but continuous onset of collective coherence.
	In contrast, few-level quantum star networks exhibit fundamentally different behavior. 
	For identical oscillators with asymmetric dissipation, 
	the system shows quasi-explosive synchronization in the weak-coupling regime, 
	which transitions into remote synchronization as the coupling strength increases.
	These two regimes originate from the competition 
	between $1\!:\!1$ and $2\!:\!1$ phase locking 
	between the hub and the leaf oscillators, 
	which are governed by the first- and second-order correlations 
	$\langle S_\mathrm{hub}^+ S_\mathrm{leaf}^- \rangle$ 
	and $\langle (S_\mathrm{hub}^+ S_\mathrm{leaf}^-)^2 \rangle$, respectively. 
	When the first-order correlation dominates, 
	the system exhibits remote synchronization, 
	whereas the second-order correlation gives rise to the interference-blockade regime. 
	We further investigate the configuration in which the hub differs from the otherwise identical leaf oscillators. 
	For sufficiently large detuning between the hub and the leaves, 
	the system first exhibits remote synchronization in the weak-coupling regime, 
	and transitions to quasi-explosive synchronization as the coupling strength increases, 
	consistent with the behavior observed in classical star networks.
	
	We first introduce our model in Sec.~\ref{sec:model}, 
	where the system dynamics are described by a Lindblad master equation 
	and the synchronization measures are defined. 
	Section~\ref{sec:identical} presents the results for the fully symmetric configuration, 
	in which the hub and the leaf oscillators are identical. 
	In Sec.~\ref{sec:nonidentical}, we analyze the synchronization behavior 
	under frequency detuning between the hub and the leaf oscillators 
	and unequal dissipation rates.
	Finally, Sec.~\ref{sec:conclusion} summarizes our main findings 
	and outlines potential directions for future research.

	\begin{figure}[t]
		\begin{overpic}[width=8.6cm]
			{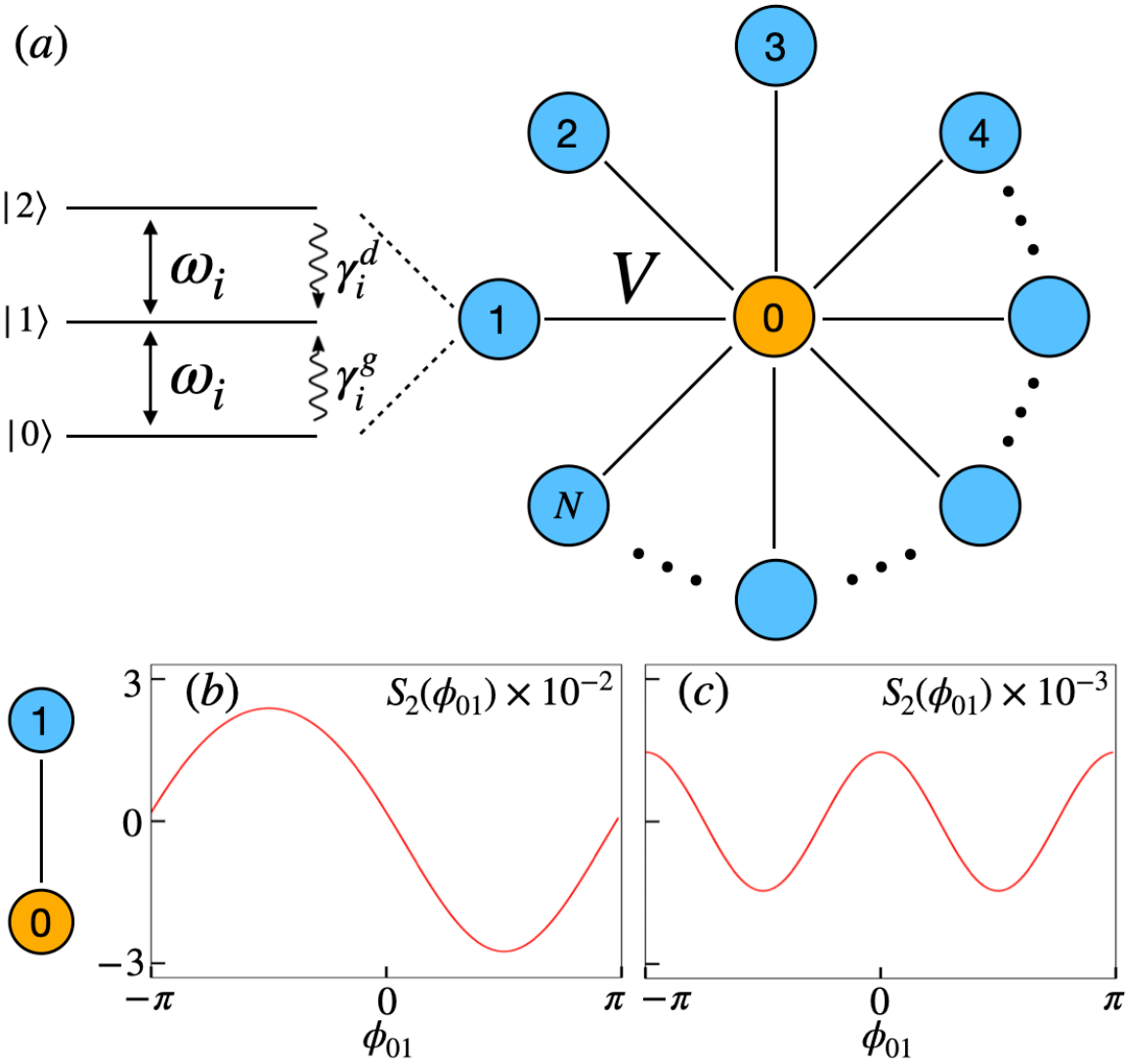}
		\end{overpic}
		\caption{(a) Schematic of the star network. 
	The central oscillator $0$ represents the hub, 
	and the outer oscillators $1$-$N$ represent the leaves. 
	Each leaf is coupled to the hub with coupling strength $V$. 
	Every oscillator consists of three spin-1 states $\ket{0}$, $\ket{1}$, and $\ket{2}$, 
	with transition frequency $\omega_i$. 
	Dissipation occurs toward the intermediate state $\ket{1}$, 
	characterized by the gain and damping rates $\gamma_i^{g}$ and $\gamma_i^{d}$, respectively.
	(b), (c) Phase distributions $S_2(\phi_{01})$ for two coupled spin-1 particles, 
	illustrating the two distinct phase-locking regimes. 
	(b) $1\!:\!1$ phase locking for 
	$\gamma_0^d = \gamma_1^g = 0.1\,\gamma_0^g = 0.1\,\gamma_1^d$. 
	(c) $2\!:\!1$ phase locking for 
	$\gamma_0^g = \gamma_1^g = 0.1\,\gamma_0^d = 0.1\,\gamma_1^d$. 
	In both cases, the coupling strength is $V = 0.05\,\gamma_1^d$.
		}
		\label{fig1}
	\end{figure}
	
	\section{Model and Theoretical Framework}\label{sec:model}
	We model the system as a star network composed of a central hub and $N$ identical leaf oscillators, as schematically shown in \cref{fig1}(a).
	Each spin-1 particle is subject to incoherent gain and damping processes, 
	in which both the upper and lower states decay into the intermediate state, 
	leading to a stable limit cycle governed by Lindbladian dynamics~\cite{roulet2018synchronizing}.
	The leaf oscillators share a common frequency \(\omega_z\); in a rotating frame at \(\omega_z\), the full system dynamics can thus be expressed by the following master equation
	\begin{align}
		\frac{\d}{\d t}\rho &= \mathcal{L}(\rho)= -\ii[H,\rho] + \sum_{j=0}^{N} \mathcal{L}_j(\rho)\,.\label{eq:EOM1}\\
		H &=  \Delta S^z_0 + \sum_{j=1}^{N} V (S^+_0 S^-_j + \text{H.c.})\,,\nonumber\\
		\mathcal{L}_j(\rho) &= \gamma^g_j\mathcal{D}[|1\rangle\langle 0|_j](\rho) + \gamma^d_j\mathcal{D}[|1\rangle\langle 2|_j](\rho)\,.\nonumber
	\end{align}
	where\(\Delta = (\omega_0-\omega_z)\) is the detuning between the hub and the leaf oscillators.
	Here, the subscript \( j=0 \) labels the central hub oscillator, 
	while \( j>0 \) labels the leaf oscillators of the network. 
	We adopt the standard spin-1 operators 
	\( S^z = \dyad{1}{1} - \dyad{-1}{-1} \) 
	and 
	\( S^{\pm} = \sqrt{2}\,(\dyad{\pm 1}{0} + \dyad{0}{\mp 1}) \). 
	Dissipation is described by the Lindblad superoperator 
	\( \mathcal{D}[L](\rho) = L\rho L^\dagger - \tfrac{1}{2}(L^\dagger L\rho + \rho L^\dagger L) \), 
	which accounts for both gain and damping processes. 
	The hub is characterized by individual rates \(\gamma_0^{g}\) and \(\gamma_0^{d}\), 
	whereas all leaf oscillators share identical rates 
	\(\gamma_N^{g}\) and \(\gamma_N^{d}\). 

	For completeness, the derivation of the effective master equation starting from the laboratory frame is detailed in Appendix~~\ref{AppendixA}.
	Applying the unitary transformation \(U(t)=\exp\big(\ii\omega_z t\sum_{j=0}^{N}S_j^z\big)\) to the lab-frame dynamics yields the Hamiltonian in Eq.~(\ref{eq:EOM1}).
	The Lindblad dissipators retain their form in the rotating frame, as the transformation induces only a global phase factor on the jump operators, which cancels out within the dissipation superoperator $\mathcal{D}[L]$.

	In previous work~\cite{roulet2018quantum}, 
	a synchronization measure was introduced to quantify 
	the degree of phase locking between two spins. 
	It is defined in terms of a specific correlation function as
	\begin{align}
		S_2(\phi_{ij}\,) =& \int\limits_0^{2\pi}\dd\phi_j \int\limits_0^\pi\dd\theta_i \sin(\theta_i)\nonumber\\
		&\times\int\limits_0^\pi\dd\theta_j \sin(\theta_j) Q(\vec{\theta}, \vec{\phi},\rho)- \frac{1}{(2\pi)^N}\,,\label{eq:Sndef}
	\end{align}
	where
	\begin{align}
		Q(\vec{\theta}, \vec{\phi}, \rho) 
		&= \left(\frac{2S+1}{4\pi}\right)^2 
		\bra{\vec{\theta}, \vec{\phi}} \rho \ket{\vec{\theta}, \vec{\phi}}, \nonumber\\
		\ket{\vec{\theta}, \vec{\phi}} 
		&= \bigotimes_{k=i,j} 
		\ee^{-\ii \phi_k S^z} \ee^{-\ii \theta_k S^y} \ket{S,S}.
	\end{align}
	Here, $\phi_{ij}$ denotes the relative phase between oscillators $i$ and $j$, 
	$Q(\vec{\theta}, \vec{\phi}, \rho)$ is the Husimi $Q$ function representing 
	the phase-space distribution of the density matrix $\rho$, 
	and $\ket{S,S}$ is the extremal spin-$S$ state. 
	This measure characterizes the distribution of relative phase, 
	as derived from the projection of the density matrix 
	onto the Bloch sphere via spin coherent states $\ket{\vec{\theta}, \vec{\phi}}$. 
	A key advantage of this measure is that it isolates correlations 
	arising purely from relative phase. 
	In practice, all contributions unrelated to the phase difference $\phi_{ij}$ 
	are removed through averaging over the remaining spin orientations, 
	so that the measure captures only the correlations associated 
	with relative-phase locking, 
	thus establishing a direct correspondence 
	with the phase distributions of classical oscillators.
	Specifically, for two spin-1 particles, the synchronization measure \(S_2\) takes the form
	\begin{align}
		S_2(\phi_{ij})
		& = \frac{9\pi}{256}|\left\langle S^+_i S^-_j\right\rangle|\cos(\phi_{ij}+\phi^{(1)})\,\nonumber\\
		&\;\; +\frac{1}{16\pi}|\left\langle(S^+_i S^-_j)^2\right\rangle|\cos(2\phi_{ij}+\phi^{(2)})\,\nonumber\\
		&= S_2^{(1)}(\phi_{ij}) + S_2^{(2)}(\phi_{ij})\,.
	\end{align}\label{eq:S2def}
	where $\phi_{ij}=\phi_i-\phi_j$ is the relative phase of two oscillators $i$ and $j$, and\(\phi^{(n)}=\arg{\left\langle (S^+_i S^-_j)^{(n)}\right\rangle}\) denotes the phase angle of the \(n\)-th order correlation function.
	This form contains both first- and second-harmonic phase-locking components, corresponding respectively to 1:1 synchronization and 2:1 blockade.
	
	\begin{figure}[t]
	\includegraphics[width=8.6cm]{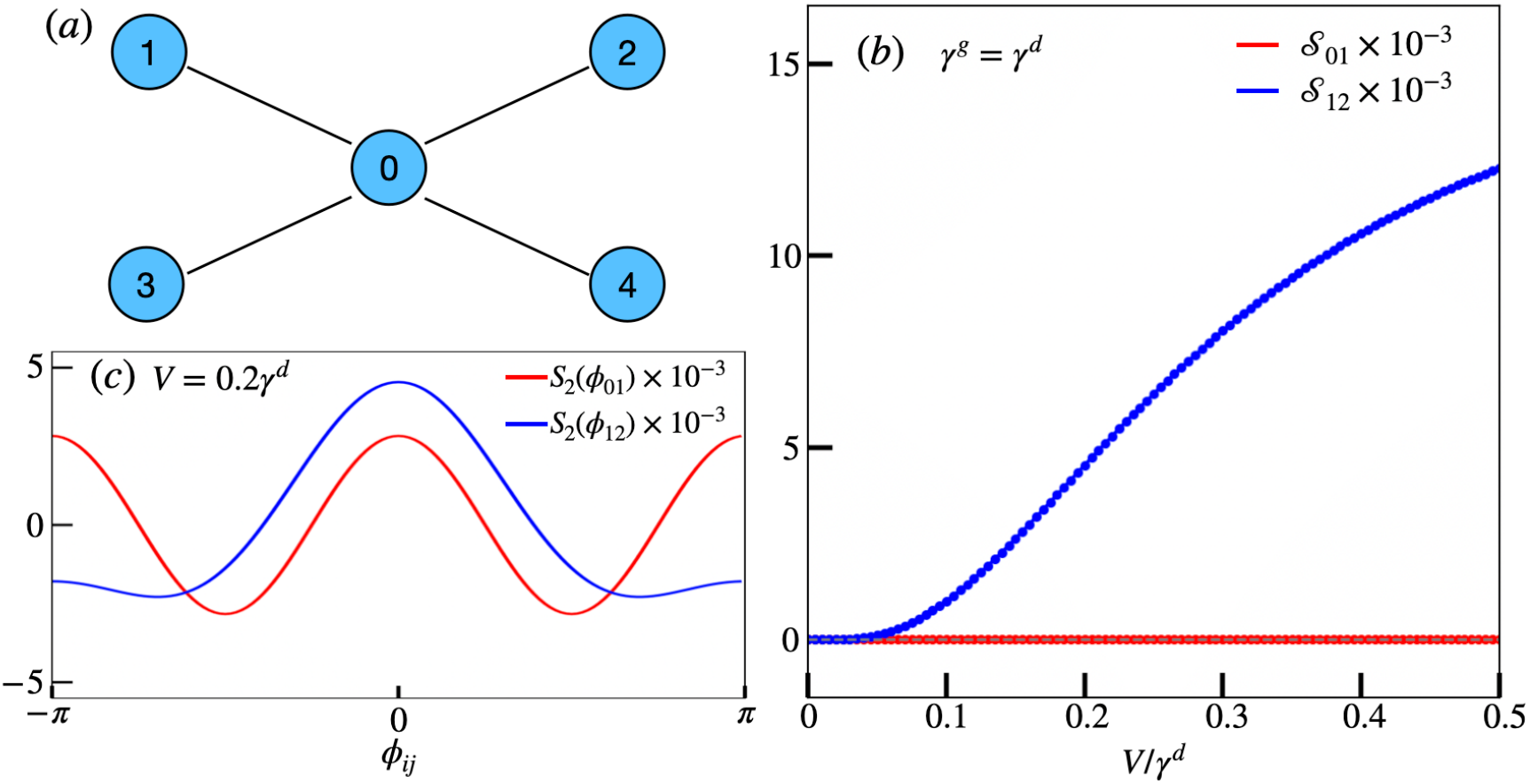}
	\caption{(a) Schematic of the star network with identical hub and leaf oscillators ($N=4$). 
	(b) Effective synchronization measures 
	$\mathcal{S}_{01}$ (hub-leaf) and $\mathcal{S}_{12}$ (leaf-leaf) 
	under symmetric gain and damping rates 
	($\gamma^{g} = \gamma^{d}$). 
	$\mathcal{S}_{01}$ remains zero, 
	whereas $\mathcal{S}_{12}$ increases monotonically with the coupling strength. 
	(c) Phase distributions $S_2(\phi_{01})$ and $S_2(\phi_{12})$ 
	at $V = 0.2\,\gamma^{d}$ corresponding to panel~(b). 
	$S_2(\phi_{01})$ exhibits two maxima, 
	while $S_2(\phi_{12})$ shows a single maximum.
		}
	\label{fig2}
	\end{figure}

	\cref{fig1}(b)(c) display the phase distributions $S_2(\phi_{01})$ 
	for two resonant spin-1 particles. 
	In \cref{fig1}(b), the first-order component $S_2^{(1)}(\phi_{01})$ dominates, 
	resulting in a single pronounced peak of $S_2(\phi_{01})$. 
	The relative phase between the two spins is therefore aligned along one direction, 
	signifying $1\!:\!1$ phase locking and synchronized motion. 
	In contrast, when the dissipation rates satisfy 
	$\gamma_0^{d} + \gamma_1^{g} = \gamma_0^{g} + \gamma_1^{d}$, 
	the first-order contribution vanishes, 
	and the phase distribution becomes entirely governed by the second-order term 
	$S_2^{(2)}(\phi_{01})$, as shown in \cref{fig1}(c). 
	It follows a $\cos(2\phi_{01})$ dependence, 
	with the relative phase adopting two equivalent orientations, 
	characteristic of $2\!:\!1$ phase locking 
	and indicative of the interference-blockade regime.
	
	To quantify the overall synchronization strength, 
	we define the \emph{effective synchronization measure} \(\mathcal{S}_{ij}\)
	as the difference between the largest and the second-largest positive peaks 
	of the phase distribution $S_2(\phi)$.
	\begin{equation}
	\label{eq:SyncContrast}
	\mathcal{S}_{ij}
	= S_2^{(\mathrm{max})}(\phi_{ij}) - S_2^{(\mathrm{2nd\,max})}(\phi_{ij}).
	\end{equation}
	If the corresponding positive peak is absent, its value is taken to be zero.
	When $\mathcal{S}_{ij}=0$, the phase distribution exhibits a two-peak structure 
	characteristic of the interference-blockade regime. 
	For $\mathcal{S}_{ij}>0$, the $1\!:\!1$ phase locking becomes dominant, 
	indicating the onset of coherent synchronization between oscillators. 
	Larger values of $\mathcal{S}_{ij}$ correspond to stronger synchronization, 
	implying that the main peak of $S_2(\phi_{ij})$ becomes more pronounced 
	and concentrated in phase.

	In the star network considered here, 
	the leaf oscillators are identical and therefore equivalent under permutation symmetry. 
	Accordingly, the synchronization properties of the entire network 
	can be fully characterized by examining a single representative pair in each category. 
	Specifically, the pair $(0,1)$ is used to quantify the synchronization 
	between the hub and the leaves, 
	whereas the pair $(1,2)$ represents the synchronization 
	between two leaves.
	
	\section{Synchronization Transmission in Identical Star Networks}\label{sec:identical}

	In this section, we consider the case where the hub and the leaf oscillators are identical, 
	forming a star network with $N=4$ leaves[see \cref{fig2}(a)]. 
	In this configuration, all oscillators share the same frequency, $\omega_0=\omega_z$, 
	and identical gain and damping rates, $\gamma_0^g=\gamma_N^g=\gamma^g$ 
	and $\gamma_0^d=\gamma_N^d=\gamma^d$. 

	We next evaluate the synchronization between the hub and a leaf, 
	and between two leaves, 
	based on the phase distributions $S_2(\phi_{ij})$ 
	and the corresponding effective synchronization measure $\mathcal{S}$. 
	To illustrate the role of dissipation asymmetry, 
	we consider different cases of gain and damping rates, 
	namely $\gamma^g = \gamma^d$, $\gamma^g \neq\,\gamma^d$. 

	In the first case of $\gamma^g = \gamma^d$, 
	the effective synchronization measures between the hub and a leaf ($\mathcal{S}_{01}$) 
	and between two leaves ($\mathcal{S}_{12}$) 
	as functions of the coupling strength are shown in \cref{fig2}(b). 
	When the gain and damping rates of each spin are identical, 
	$\mathcal{S}_{01}$ remains zero, indicating that the hub-leaf pairs 
	stay in the interference-blockade regime without establishing direct synchronization. 
	In contrast, $\mathcal{S}_{12}$ increases monotonically with the coupling strength, 
	demonstrating that synchronization between the leaves 
	is mediated through the hub's interference-blockade channel. 
	This configuration therefore realizes remote synchronization, 
	where the leaf oscillators become mutually synchronized 
	via a blockaded hub that itself remains unsynchronized.
	To substantiate this mediated-locking mechanism, 
	\cref{fig2}(c) displays the phase distributions $S_2(\phi_{01})$ and $S_2(\phi_{12})$
	at $V = 0.2\,\gamma^d$. 
	As shown, $S_2(\phi_{01})$ exhibits two symmetric peaks of equal height, 
	characteristic of a $2\!:\!1$ phase locking in the interference-blockade regime, 
	whereas $S_2(\phi_{12})$ displays a single dominant peak,
	corresponding to a $1\!:\!1$ phase locking and synchronization 
	between the leaves.	

	\begin{figure}[t]
		\includegraphics[width=8.6cm]{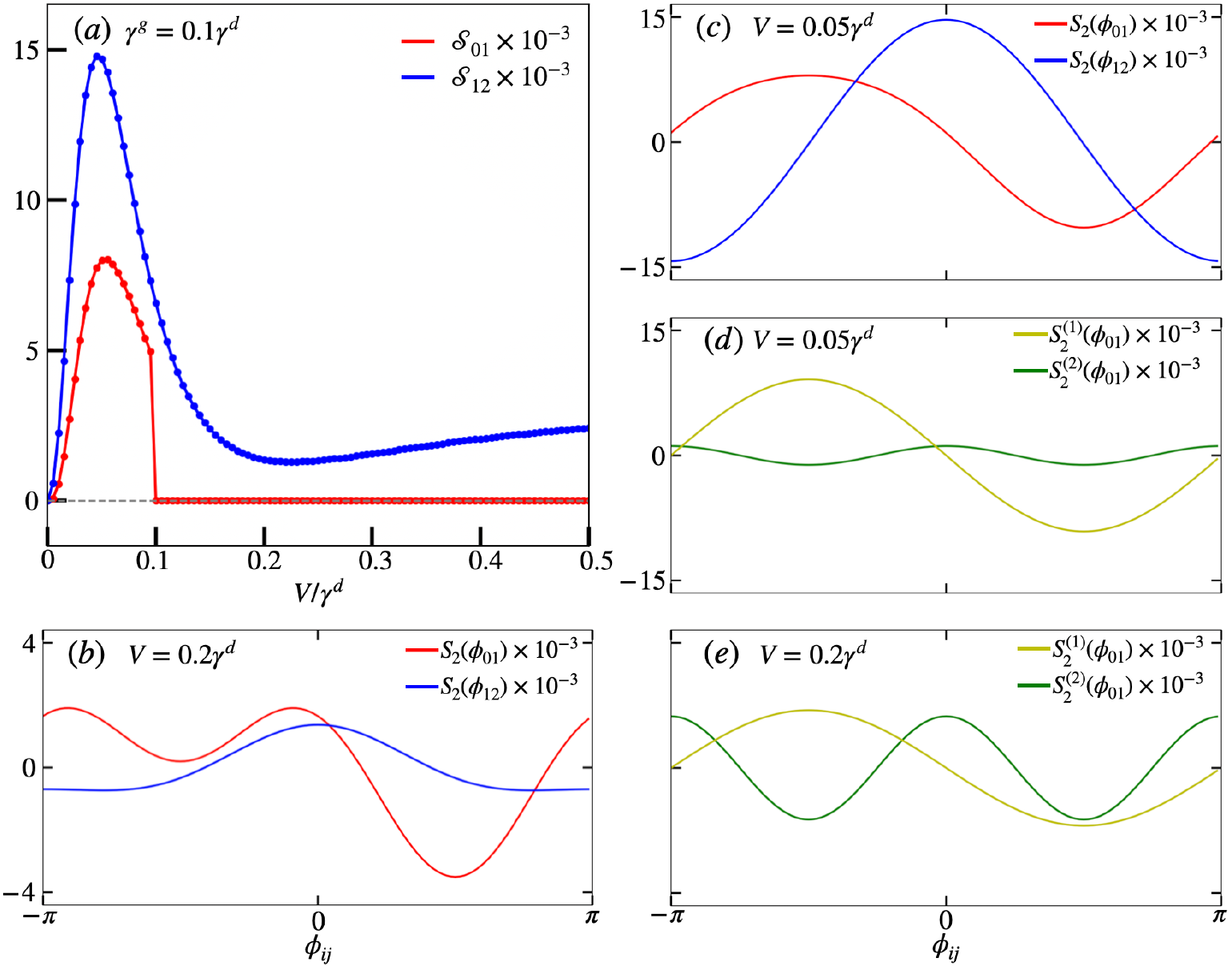}
		\caption{(a) Effective synchronization measures 
		$\mathcal{S}_{01}$ (hub-leaf) and $\mathcal{S}_{12}$ (leaf-leaf) 
		for gain and damping rates $\gamma^{g} = 0.1\,\gamma^{d}$. 
		Both measures exhibit nonmonotonic behavior: 
		$\mathcal{S}_{01}$ first increases and then decreases to zero, 
		while $\mathcal{S}_{12}$ shows a similar trend, 
		reaching its maximum simultaneously with $\mathcal{S}_{01}$ 
		before decreasing and increasing again at stronger coupling. 
		(b), (c) Phase distributions $S_2(\phi_{01})$ and $S_2(\phi_{12})$ 
		for $V = 0.2\,\gamma^{d}$ and $V = 0.05\,\gamma^{d}$, respectively, 
		under the same conditions as in panel~(a). 
		In both cases, $S_2(\phi_{12})$ displays a single maximum, 
		whereas $S_2(\phi_{01})$ exhibits two maxima at $V = 0.2\,\gamma^{d}$ 
		and a single maximum at $V = 0.05\,\gamma^{d}$. 
		(d), (e) First- and second-order contributions 
		$S_2^{(1)}(\phi_{01})$ and $S_2^{(2)}(\phi_{01})$ 
		for $V = 0.05\,\gamma^{d}$ and $V = 0.2\,\gamma^{d}$, respectively. 
		At $V = 0.05\,\gamma^{d}$, the first-order term $S_2^{(1)}(\phi_{01})$ dominates, 
		whereas at $V = 0.2\,\gamma^{d}$, 
		the first- and second-order components become comparable.
		}
		\label{fig3}
	\end{figure}

	We now examine the regime of asymmetric dissipation, 
	where the gain and damping rates are no longer equal. 
	We first focus on the case $\gamma^g < \gamma^d$. 
	For a two-oscillator system ($N=1$), this configuration still fulfills 
	the synchronization-blockade condition. 
	The key question is whether such a blockade persists 
	once the system is extended to a star network, 
	and whether synchronization transmission can still occur 
	through the interference-blockade channel. 
	\cref{fig3}(a) displays the effective synchronization measures 
	$\mathcal{S}_{01}$ and $\mathcal{S}_{12}$ for $\gamma^g = 0.1\,\gamma^d$. 
	In contrast to the symmetric case of $\gamma^g = \gamma^d$, 
	$\mathcal{S}_{12}$ exhibits a nonmonotonic dependence on the coupling strength—%
	it first increases, then decreases, and subsequently rises again. 
	At the same time, $\mathcal{S}_{01}$ follows a similar trend, 
	increasing together with $\mathcal{S}_{12}$ and reaching its maximum at the same coupling strength, 
	after which it rapidly drops to zero.
	
	These results reveal two distinct synchronization mechanisms 
	that emerge in the weak- and strong-coupling regimes, respectively. 
	At weak coupling, as the coupling strength increases, 
	the entire network rapidly develops global synchronization, 
	resembling a quasi-explosive synchronization transition. 
	When the coupling exceeds a certain threshold, however, 
	the hub abruptly desynchronizes and enters the interference-blockade regime, 
	while the leaves remain mutually synchronized, 
	giving rise to remote synchronization.

	\begin{figure}[t]
		\includegraphics[width=7.6cm]{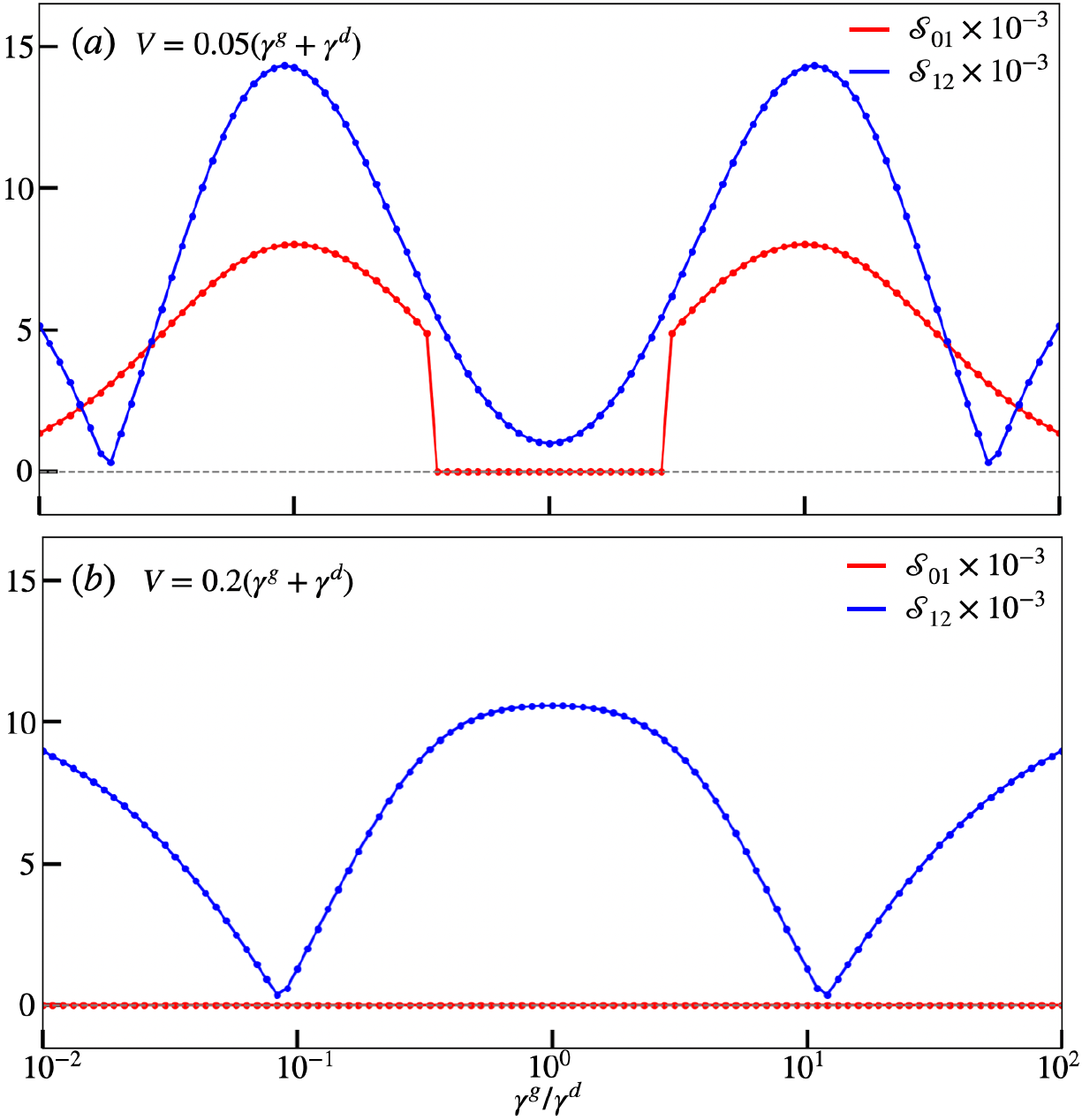}
		\caption{Effective synchronization measures 
		$\mathcal{S}_{01}$ and $\mathcal{S}_{12}$ 
		evaluated at two coupling strengths: 
		(a) $V = 0.05\,(\gamma^{g} + \gamma^{d})$; 
		(b) $V = 0.2\,(\gamma^{g} + \gamma^{d})$.
		}
		\label{fig4}
	\end{figure}

	To further investigate this nonmonotonic behavior,
	we compute the relative phase distributions $S_2(\phi_{01})$ and $S_2(\phi_{12})$ 
	for two representative coupling strengths, $V = 0.05\,\gamma^d$ and $V = 0.2\,\gamma^d$, 
	as shown in \cref{fig3}(b)(c), respectively. 
	These two coupling values correspond to the regions in \cref{fig3}(a) 
	where $\mathcal{S}_{01}$ is finite and where it vanishes, 
	corresponding to the two synchronization mechanisms discussed above.
	In the weak-coupling regime, $S_2(\phi_{01})$ exhibits a single dominant peak, 
	indicating that the hub and the leaves are phase locked. 
	As the coupling strength increases, 
	this single peak gradually splits into two peaks of equal height, 
	marking the emergence of the interference-blockade regime 
	in which the hub is desynchronized.

	To understand this transition, 
	we further decompose $S_2(\phi_{01})$ into its first- and second-order contributions, 
	$S_2^{(1)}(\phi_{01})$ and $S_2^{(2)}(\phi_{01})$, as shown in \cref{fig3}(d)(e). 
	It is evident that these two components respond differently to the coupling strength. 
	In the weak-coupling regime, the peak amplitude of $S_2^{(1)}(\phi_{01})$ 
	is much larger than that of $S_2^{(2)}(\phi_{01})$, 
	so that the total $S_2(\phi_{01})$ exhibits a single-peak structure[red line in \cref{fig3}(c)].
	As the coupling strength increases, however, 
	the first-order component $S_2^{(1)}(\phi_{01})$ decreases 
	and becomes comparable in magnitude to the second-order contribution $S_2^{(2)}(\phi_{01})$. 
	The combination of these two components results in a double-peak structure 
	in the total distribution $S_2(\phi_{01})$ [red line in \cref{fig3}(b)].
	Therefore, the two distinct synchronization behaviors observed in the star network 
	arise from the competition between the first- and second-order contributions, 
	$S_2^{(1)}(\phi_{01})$ and $S_2^{(2)}(\phi_{01})$. 
	This competition occurs only when the single-peak of the first-order term 
	and the double peaks of the second-order term 
	do not coincide in phase, 
	that is, when the phase shifts satisfy 
	$\phi^{(1)}_{01} \neq \phi^{(2)}_{01} \,(\mathrm{mod}\,\pi)$. 
	If the two contributions were phase aligned, 
	their superposition would yield merely a constructive enhancement, 
	producing a single dominant peak instead of a split structure.
	This nonmonotonic synchronization behavior, 
	absent in classical star networks, 
	arises from the competition between dissipation and coupling 
	in the quantum open system.

	\begin{figure}[t]
		\includegraphics[width=8.6cm]{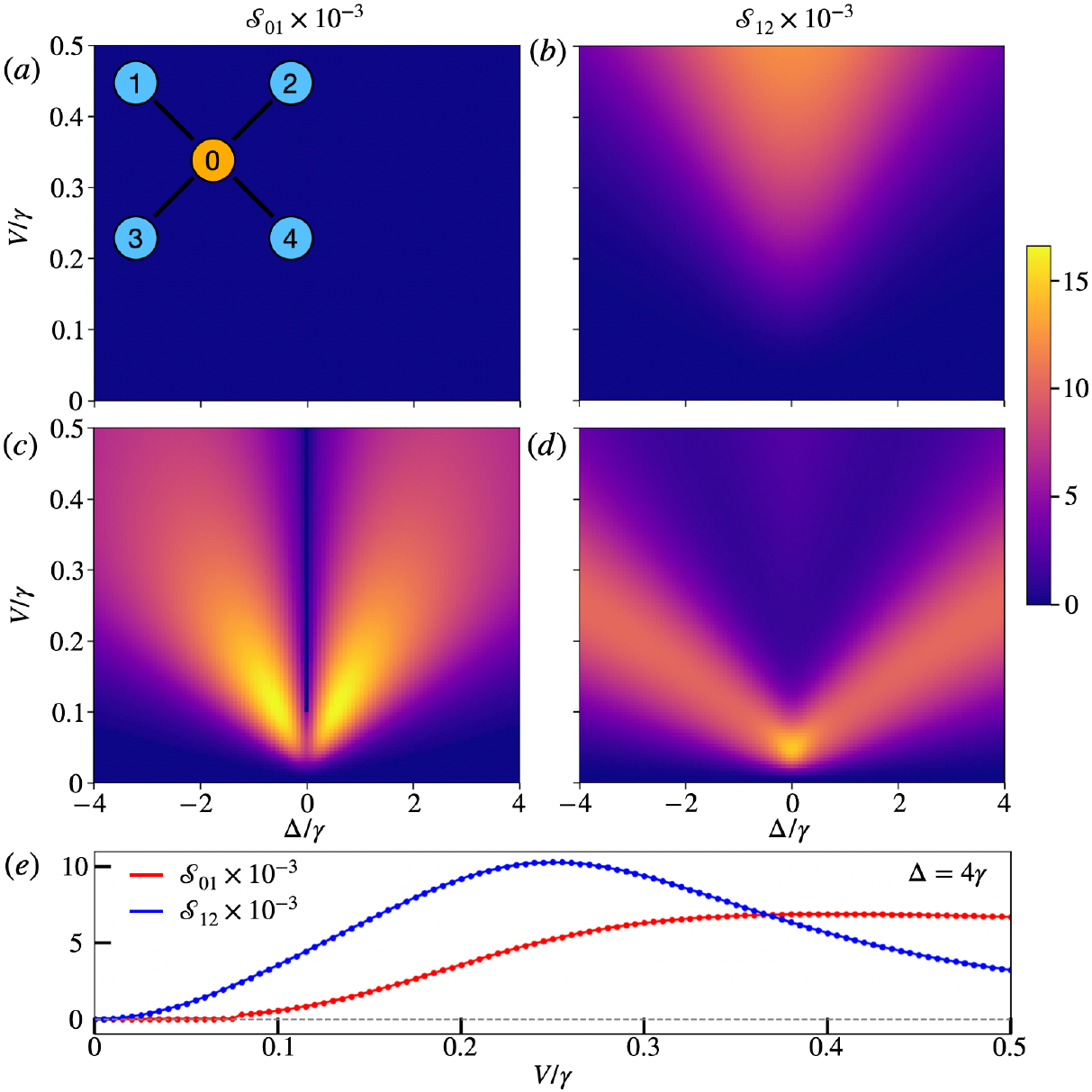}
		\caption{(a)–(e) Effective synchronization measures 
		$\mathcal{S}_{01}$  and $\mathcal{S}_{12}$ 
		for a star network with one detuned hub and four identical leaves ($N=4$). 
		The inset in panel~(a) illustrates the network configuration. 
		(a), (b) Symmetric dissipation, 
		$\gamma_0^g = \gamma_N^g = \gamma_0^d = \gamma_N^d = \gamma$. 
		$\mathcal{S}_{01}$ remains zero for all detunings and coupling strengths, 
		whereas $\mathcal{S}_{12}$ decreases with detuning and increases with coupling. 
		(c)-(e) Asymmetric dissipation, 
		$\gamma_0^g = \gamma_N^g = 0.1\,\gamma_0^d = 0.1\,\gamma_N^d = 0.1\,\gamma$. 
		Under resonance, $\mathcal{S}_{01}$ first increases and then drops to zero, 
		while under detuning it decreases to a finite value. 
		$\mathcal{S}_{12}$ exhibits a similar nonmonotonic behavior, 
		increasing, then decreasing, and gradually rising again near resonance.
		(e) Same asymmetric dissipation as in panels~(c),(d), 
		with fixed detuning $\Delta = 4\,\gamma$.
		}
		\label{fig5}
	\end{figure}

	Finally, we investigate the synchronization behavior of the oscillators 
	as a function of the ratio between the gain and damping rates. 
	The effective synchronization measures $\mathcal{S}_{01}$ and $\mathcal{S}_{12}$ 
	are computed under various dissipation conditions, as shown in \cref{fig4}.
	In the weak-coupling regime, \cref{fig4}(a) shows that 
	synchronization blockade between the hub and the leaves 
	appears only in the vicinity of the symmetric point $\gamma^g = \gamma^d$. 
	As the coupling strength increases, the blockade region progressively broadens, 
	as illustrated in \cref{fig4}(b). 
	Thus, whether the star network exhibits remote synchronization 
	or transitions to global synchronization 
	is determined by the interplay between dissipation and coupling, 
	consistent with our previous analysis.

	\section{Synchronization Transmission with a Detuned and Dissipation-Imbalanced Hub}\label{sec:nonidentical}
	We now consider a star network in which the leaf nodes remain identical, 
	while the hub differs in both its detuning and dissipation rates. 
	We also fix the number of leaf nodes to $N=4$, as illustrated in the inset of \cref{fig5}(a). 
	Following the convention of Sec.~\ref{sec:model}, 
	the detuning is defined as $\Delta = \omega_0 - \omega_z$, 
	and the gain and damping rates of the hub and the leaves are denoted by 
	$\gamma_0^{g}$, $\gamma_0^{d}$, $\gamma_N^{g}$, and $\gamma_N^{d}$, respectively.
	
	We first investigate the influence of detuning on the effective synchronization measures 
	$\mathcal{S}_{01}$ and $\mathcal{S}_{12}$, 
	using the same dissipation parameters as in Sec.~\ref{sec:identical}. 
	The corresponding results are presented in \cref{fig5}. 
	Under symmetric dissipation 
	($\gamma_0^g = \gamma_N^g = \gamma_0^d = \gamma_N^d = \gamma^d$), 
	$\mathcal{S}_{01}$ remains zero even in the presence of detuning, 
	whereas $\mathcal{S}_{12}$ increases with the coupling strength 
	and decreases with larger detuning. 
	This dependence forms a structure characteristic of the classical Arnold tongue, 
	as shown in \cref{fig5}(b). 
	These findings demonstrate that, in quantum remote synchronization, 
	the response of leaf-leaf synchronization to detuning and coupling 
	mirrors that of two directly coupled spins.

	\cref{fig5}(c)(d) show the results obtained for the dissipation parameters 
	$\gamma_0^g = \gamma_N^g = 0.1\,\gamma^d$ and 
	$\gamma_0^d = \gamma_N^d = \gamma^d$. 
	When the hub and the leaf nodes are resonant, 
	the system exhibits synchronization in the weak-coupling regime 
	and transitions to a blockade behavior as the coupling strength increases. 
	In contrast, when a detuning is introduced between the hub and the leaves, 
	the growth of synchronization between the hub and a leaf 
	and that among the leaves becomes markedly different. 
	Unlike the resonant case, $\mathcal{S}_{01}$ does not increase immediately with the onset of coupling; 
	instead, its growth is delayed, and the coupling strength required to initiate synchronization 
	increases progressively with larger detuning. 
	To further illustrate this behavior, 
	\cref{fig5}(e) presents $\mathcal{S}_{01}$ and $\mathcal{S}_{12}$ for a detuning of $\Delta = 4\,\gamma$. 
	In the weak-coupling regime, $\mathcal{S}_{01}$ remains zero, indicating that the hub-leaf pairs 
	are in the blockade regime, whereas $\mathcal{S}_{12}$ increases with coupling strength, 
	signaling the emergence of remote synchronization. 
	As the coupling is further enhanced, both $\mathcal{S}_{01}$ and $\mathcal{S}_{12}$ increase sharply, 
	marking the onset of quasi-explosive synchronization. 
	Such behavior at large detuning is consistent with that observed in classical star networks.

	\begin{figure}[t]
		\includegraphics[width=8.6cm]{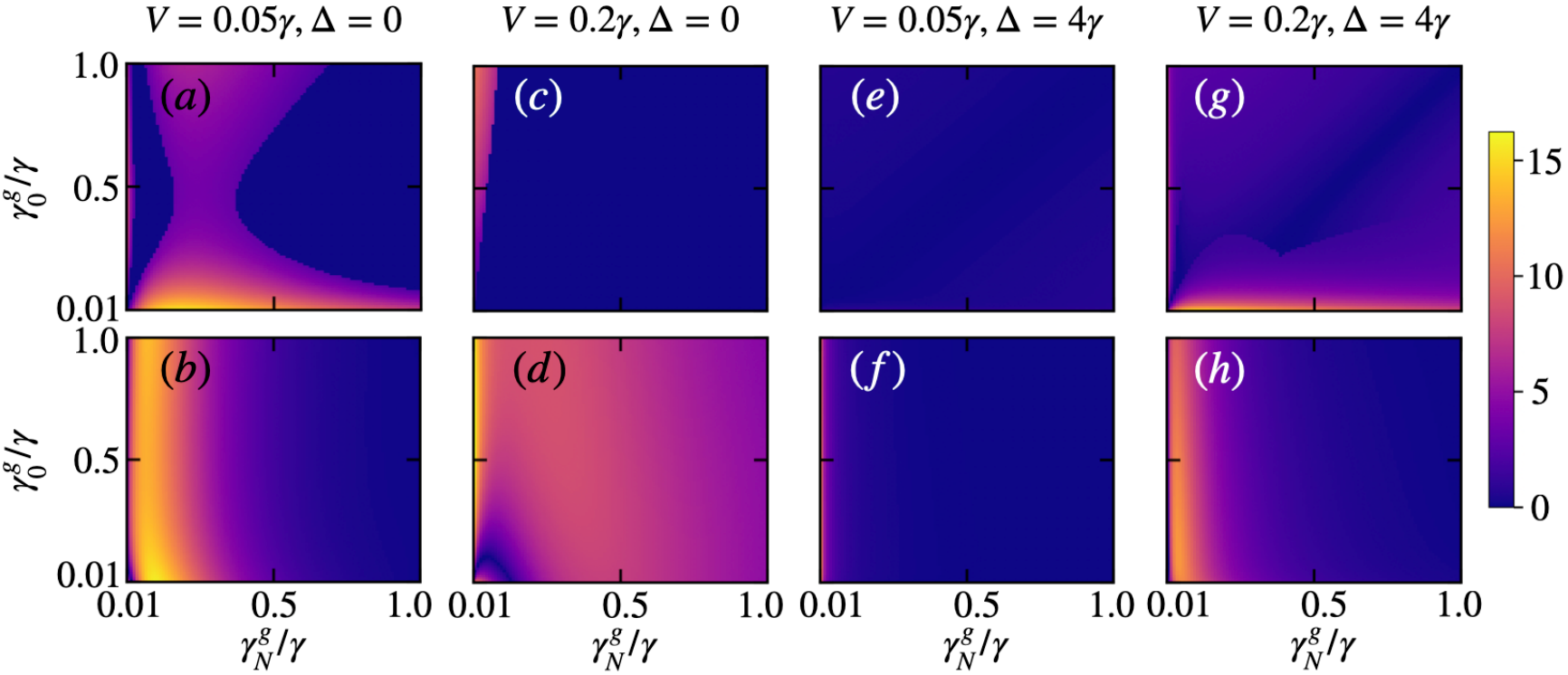}
		\caption{(a)–(h) Effective synchronization measures 
		$\mathcal{S}_{01}$ and $\mathcal{S}_{12}$ 
		for varying dissipation rates. 
		The top row [(a), (c), (e), (g)] shows $\mathcal{S}_{01}$ 
		(hub-leaf synchronization), 
		and the bottom row [(b), (d), (f), (h)] shows $\mathcal{S}_{12}$ 
		(leaf-leaf synchronization). 
		The damping rates are fixed at 
		$\gamma_0^d = \gamma_N^d = \gamma$.
		}
		\label{fig6}
	\end{figure}
	
	The contrasting synchronization trends in Fig.~\ref{fig3} and Fig.~\ref{fig5} arise from the competition and phase relationship between first-order and second-order correlations.
	In the resonant regime (Fig.~\ref{fig3} and Fig.~\ref{fig4}), the first-order correlation $\langle S_{0}^{+}S_{j}^{-}\rangle$ is progressively counteracted by the second-order correlation $\langle (S_{0}^{+}S_{j}^{-})^2 \rangle$ due to destructive interference.
	As the coupling strength increases, this strictly competitive relationship drives the direct hub-leaf synchronization $\mathcal{S}_{01}$ to zero (interference blockade).
	In contrast, detuning (Fig.~\ref{fig5}) modifies the relative phase between these two correlations. They are no longer locked into a strictly destructive relationship; instead, their interplay shifts from competition to cooperation depending on the detuning magnitude.
	Consequently, the effective strength of the first-order correlation is modulated—being either partially weakened or constructively enhanced—allowing $\mathcal{S}_{01}$ to vary significantly and potentially exceed $\mathcal{S}_{12}$ in specific parameter regimes.

	Finally, we present the effective synchronization measures for different dissipation rates 
	of the hub and the leaf oscillators, as shown in \cref{fig6}. 
	Under both resonant and detuned conditions, 
	the blockade regions exhibit distinct responses to variations in the gain rates. 
	In the resonant case, the blockade region between the leaves is relatively narrow, 
	and strong coupling facilitates robust remote synchronization. 
	At weak coupling, quasi-explosive synchronization emerges 
	when the gain rates of both the hub and the leaves are small. 
	In contrast, under detuned conditions, 
	the parameter region that supports remote synchronization becomes substantially reduced, 
	and remote synchronization appears only when the gain rates of the leaves are sufficiently small.

	\section{Conclusion}\label{sec:conclusion}
	We have analyzed the synchronization transmission phenomena in both identical star networks 
	and in networks where the leaves are identical but the hub differs. 
	Under different dissipation rates, we observe both remote synchronization and quasi-explosive synchronization, 
	phenomena also found in classical systems. 
	However, the mechanisms and conditions for their emergence are specific to quantum systems 
	and differ from those in classical star networks.

	In the configuration where the hub and the leaf oscillators are identical, 
	the synchronization blockade between them occurs when the first-order phase correlation no longer dominates. 
	For equal gain and damping rates, the first-order contribution in the hub-leaf phase distribution vanishes, 
	leaving only the second-order term that gives rise to the $2\!:\!1$ phase-locking blockade. 
	In contrast, the leaves exhibit a single-peak phase distribution corresponding to $1\!:\!1$ phase locking, 
	and the synchronization strength increases monotonically with coupling, realizing remote synchronization. 
	Introducing an imbalance between the gain and damping rates restores the first-order contribution 
	in the hub-leaf synchronization, 
	which dominates in the weak-coupling regime and produces quasi-explosive behavior, 
	while stronger coupling leads to a $2\!:\!1$ blockade and remote synchronization through the hub.  

	For networks where the leaves remain identical but the hub differs, 
	the synchronization behavior depends sensitively on both dissipation and detuning. 
	Under symmetric dissipation, the hub-leaf blockade persists even with finite detuning, 
	and the leaf-leaf synchronization follows an Arnold-tongue-like dependence, 
	providing favorable conditions for remote synchronization. 
	When dissipation imbalance is introduced, 
	the synchronization dynamics exhibit opposite trends under resonant and detuned conditions. 
	In the resonant case, the system shows quasi-explosive synchronization at weak coupling 
	and transitions to remote synchronization as the coupling strength increases. 
	In contrast, under detuning, remote synchronization emerges in the weak-coupling regime, 
	followed by quasi-explosive synchronization at stronger coupling. 
	Such detuning-induced behavior is consistent with that observed in classical star networks.

	The mediated transmission of quantum synchronization revealed in this work 
	exhibits a rich variety of synchronization behaviors, 
	including but not limited to remote and quasi-explosive synchronization. 
	We expect that these findings will provide a foundation for further studies 
	of quantum synchronization transmission in larger and more complex systems. 
	In particular, explosive synchronization may emerge in networks with more oscillators, 
	while partial or cluster-type remote synchronization could arise in networks 
	with heterogeneous coupling or dissipation structures.

	\acknowledgments
	This work was supported by the National Key Research and Development Program of China (Grant No. 2022YFA1405301 and No. 2018YFA0306502), the National Natural Science Foundation of China (Grant No. 12022405 and No. 11774426).

	
	\appendix
	\section{Derivation of the Effective Master Equation}
	\label{AppendixA}
	\setcounter{equation}{0}
	\renewcommand{\theequation}{A\arabic{equation}}

	In this appendix, we provide a detailed derivation of the effective master equation in the rotating frame (Eq.~(1) in the main text), starting from the full system dynamics in the laboratory frame.

	\subsection{Master Equation in the Laboratory Frame}
	The dynamics of the star network in the laboratory frame are governed by the Lindblad master equation:
	\begin{equation}
		\frac{\dd \rho_{\rm lab}}{\dd t} = -\ii [H_{\rm lab}, \rho_{\rm lab}] + \sum_{j=0}^N \mathcal{L}_j(\rho_{\rm lab}).
		\label{eq:lab_master}
	\end{equation}
	Here, the total Hamiltonian $H_{\rm lab}$ describes the coherent evolution of the hub (index $0$) and the $N$ leaves (indices $j=1,\dots,N$). We model the coupling between the hub and the leaves as an exchange interaction, given by:
	\begin{equation}
		H_{\rm lab} = \omega_0 S_0^z + \sum_{j=1}^N \omega_z S_j^z + V \sum_{j=1}^N \left( S_0^+ S_j^- + S_0^- S_j^+ \right),
	\end{equation}
	where $\omega_0$ and $\omega_z$ are the transition frequencies of the hub and leaves, respectively, and $V$ is the coupling strength. $S_k^\pm$ and $S_k^z$ are the standard spin-1 operators for the $k$-th oscillator.

	The incoherent dynamics are described by the Lindblad superoperators $\mathcal{L}_j$. For each oscillator $j$, the dissipation includes both gain and damping processes:
	\begin{equation}
		\mathcal{L}_j(\rho) = \gamma_j^g \mathcal{D}[|1\rangle\langle 0|_j]\rho + \gamma_j^d \mathcal{D}[|1\rangle\langle 2|_j]\rho,
	\end{equation}
	where the standard superoperator is defined as $\mathcal{D}[L]\rho = L\rho L^\dagger - \frac{1}{2}\{L^\dagger L, \rho\}$.

	\subsection{Transformation to the Rotating Frame}
	We move to a frame rotating at the frequency of the leaf oscillators, $\omega_z$. This is defined by the unitary transformation:
	\begin{equation}
		U(t) = \exp\left[ \ii \omega_z t \left( S_0^z + \sum_{j=1}^N S_j^z \right) \right].
	\end{equation}
	Let $\rho(t) = U(t) \rho_{\rm lab}(t) U^\dagger(t)$ be the density matrix in the rotating frame. Differentiating $\rho(t)$ with respect to time and substituting Eq.~(\ref{eq:lab_master}) yields the transformed master equation:
	\begin{equation}
		\frac{\dd \rho}{\dd t} = -\ii [H_{\rm eff}, \rho] + \sum_{j=0}^N \tilde{\mathcal{L}}_j(\rho).
	\end{equation}
	Below, we derive the explicit forms of the effective Hamiltonian $H_{\rm eff}$ and the transformed dissipators $\tilde{\mathcal{L}}_j$.

	\subsubsection{Transformation of the Hamiltonian}
	The effective Hamiltonian in the rotating frame is given by:
	\begin{equation}
		H_{\rm eff} = U(t) H_{\rm lab} U^\dagger(t) + \ii \dot{U}(t) U^\dagger(t).
	\end{equation}
	The second term (inertial term) accounts for the time-dependence of the reference frame:
	\begin{equation}
		\ii \dot{U}(t) U^\dagger(t) = -\omega_z \left( S_0^z + \sum_{j=1}^N S_j^z \right).
	\end{equation}
	For the first term, we transform the operators in $H_{\rm lab}$. The diagonal operators $S_k^z$ commute with $U(t)$ and remain unchanged. The ladder operators transform as $U(t) S_k^\pm U^\dagger(t) = S_k^\pm \ee^{\pm \ii\omega_z t}$.
	Applying this to the interaction term:
	\begin{align}
		&U(t) \left( S_0^+ S_j^- + S_0^- S_j^+ \right) U^\dagger(t) \nonumber \\
		&= (S_0^+ \ee^{\ii\omega_z t})(S_j^- \ee^{-\ii\omega_z t}) + (S_0^- \ee^{-\ii\omega_z t})(S_j^+ \ee^{\ii\omega_z t}) \nonumber \\
		&= S_0^+ S_j^- + S_0^- S_j^+.
	\end{align}
	Note that the time-dependent phase factors $\ee^{\pm i\omega_z t}$ cancel out exactly because the interaction conserves the total excitation number.
	Combining the inertial term with the transformed $H_{\rm lab}$, we obtain:
	\begin{align}
		H_{\rm eff} &= \left( \omega_0 S_0^z + \sum_{j=1}^N \omega_z S_j^z \right) - \omega_z \left( S_0^z + \sum_{j=1}^N S_j^z \right) \nonumber \\
		&+ V \sum_{j=1}^N (S_0^+ S_j^- + {\rm H.c.}) \nonumber \\
		&= (\omega_0 - \omega_z) S_0^z + V \sum_{j=1}^N (S_0^+ S_j^- + {\rm H.c.}) \nonumber \\
		&= \Delta S_0^z + V \sum_{j=1}^N (S_0^+ S_j^- + {\rm H.c.}),
	\end{align}
	which matches the Hamiltonian in Eq.~(\ref{eq:EOM1}).

	\subsubsection{Transformation of the Lindblad Dissipators}
	The transformed Lindblad superoperator is given by $\tilde{\mathcal{L}}_j(\rho) = U \mathcal{L}_j(U^\dagger \rho U) U^\dagger$. Consider a generic dissipator $\mathcal{D}[L]$ with jump operator $L$. Its transformation is:
	\begin{align}
		&U \left( L (U^\dagger \rho U) L^\dagger - \frac{1}{2} \{ L^\dagger L, U^\dagger \rho U \} \right) U^\dagger \nonumber \\
		&= \tilde{L} \rho \tilde{L}^\dagger - \frac{1}{2} \{ \tilde{L}^\dagger \tilde{L}, \rho \} = \mathcal{D}[\tilde{L}]\rho,
	\end{align}
	where $\tilde{L} = U L U^\dagger$ is the transformed jump operator.
	Our jump operators are $L_g = |1\rangle\langle 0|$ and $L_d = |1\rangle\langle 2|$. Since they are eigen-operators of $S^z$ (i.e., $[S^z, L_g] = 1 \cdot L_g$ and $[S^z, L_d] = -1 \cdot L_d$), they transform simply by acquiring a phase factor:
	\begin{equation}
		\tilde{L}_g = \ee^{\ii\omega_z t} L_g, \quad \tilde{L}_d = \ee^{-\ii\omega_z t} L_d.
	\end{equation}
	Crucially, the Lindblad form is invariant under a global phase shift of the jump operator. Substituting $\tilde{L} = \ee^{\ii\phi} L$ into $\mathcal{D}[\tilde{L}]$:
	\begin{align}
		&\mathcal{D}[\ee^{\ii\phi} L]\rho = (\ee^{\ii\phi} L) \rho (\ee^{-\ii\phi} L^\dagger) - \frac{1}{2} \{ (\ee^{-\ii\phi} L^\dagger)(\ee^{\ii\phi} L), \rho \} \nonumber \\
		&= L \rho L^\dagger - \frac{1}{2} \{ L^\dagger L, \rho \} = \mathcal{D}[L]\rho.
	\end{align}
	Thus, $\tilde{\mathcal{L}}_j(\rho) = \mathcal{L}_j(\rho)$. The form of the dissipation remains identical in the rotating frame.


	\bibliographystyle{apsrev4-2}
	\bibliography{ref}

\end{document}